\documentclass[aps, prc, preprint, groupedaddress, showpacs, showkeys,nofootinbib]{revtex4-1}

\usepackage{graphicx}
\usepackage{lineno}

\begin{document}

\title{$B_c$\ meson enhancement and the momentum dependence in Pb+Pb collisions at LHC energy}


\author{Yunpeng Liu}
\email[]{liuyp06@mails.tsinghua.edu.cn}
\author{Carsten Greiner}
\author{Andriy Kostyuk}
\affiliation{Institut f\"ur Theoretische Physik, Johann Wolfgang Goethe-Universit\"at Frankfurt, Max-von-Laue-Strasse 1, D-60438 Frankfurt am Main, Germany}


\date{\today}

\begin{abstract}
     $B_c$\ meson production in Pb+Pb collisions at $\sqrt{s}=2.76\ A$\ TeV is surveyed in both a statistical coalescence model and a transport model. The nuclear modification factor $R_{AA}$\ is predicted to be between $2$\ and $18$\ in the most central collisions, which can help to confirm the regeneration mechanism. In addition, the momentum dependence is also investigated as given by the transport model. A strong suppression of the transverse momentum is found in central collisions accompanying the enhancement in yield. The spectrum and elliptic flow of $B_c$\ are also discussed.
\end{abstract}


\maketitle

\section{Introduction}
Several kinds of signals of the hot quark matter created in relativistic heavy-ion collisions are widely discussed in the literature.  One of them is based on the hard probes, for example quenching~\cite{Adcox:2001jp} of the  jets which pass through the hot medium.  Another one is focused on the low energy particles, for example the collective flow of low $p_T$\ particles~\cite{Adare:2006ti}, which carries the information on hydrodynamical properties of the hot matter at the initial stage of the fireball evolution.
\par
One more significant idea utilizes heavy quarkonia as hard probes of the fireball. The anomalous $J/\psi$\ suppression in quark-gluon plasma (QGP) was predicted theoretically~\cite{Matsui:1986dk} a quarter of a century ago.  Such a suppression was indeed observed in heavy-ion collisions~\cite{Gonin:1996wn}. It was however suggested that this suppression can be as well explained within the models that do not assume quark deconfinement~\cite{Spieles:1999kp,Geiss:1998ma,Armesto:1998rc,Kahana:1999qp,Qiu:2001at}. Moreover, it was suggested that not only dissociation of quarkonia but also the opposite process, recombination of heavy quark-antiquark pairs, can take place in the deconfined medium~\cite{Kabana:2000iu,BraunMunzinger:2000px,Gorenstein:2000ck,Gorenstein:2000nd,Thews:2000rj, Grandchamp:2001pf, Zhang:2002ug,Kostyuk:2005zd,Yan:2007zze, Zhao:2007hh}. If the regeneration of quarkonia indeed takes place, it becomes especially significant with increasing collision energy since the number of heavy quarks per collision becomes larger. When the regeneration is dominant, quarkonia can be used to detect the early fireball through its spectrum and flow like the soft probes. Even enhancement of quarkonia production has been expected~\cite{Thews:2000rj}, which could be regarded as direct evidence of the regeneration mechanism. However, such an enhancement of heavy quarkonia is not supported at RHIC energy by the statistical model~\cite{Kostyuk:2003kt, BraunMunzinger:2000px} and has never been observed at RHIC or LHC~\cite{Adare:2006ns, Adare:2008sh, Abelev:2009qaa, Abelev:2012rv, Reed:2011fr, Chatrchyan:2011pe}.
\par
How can we verify the regeneration mechanism firmly?
\par
The $B_c$\ meson was found by CDF in 1998~\cite{Abe:1998fb}. Similar to heavy quarkonia it consists of a heavy quark and a heavy antiquark. But unlike in $J/\psi$\ or $\Upsilon$ mesons, the quark and antiquark have different flavors. This has a drastic effect on the production cross section of the $B_c$\ meson in elementary hadron-hadron collisions. Indeed, creation of one $c\bar{c}$ or $b\bar{b}$ in an elementary collision is sufficient for production of, respectively, $J/\psi$\ or $\Upsilon$. In contrast, production of a $B_c$\ meson in  hadron-hadron collisions requires creation of at least two heavy quark pairs, $c\bar{c}$ and $b\bar{b}$, in the same collision. 

Unlike in elementary hadron-hadron reactions, $B_c$\ meson production in nucleus-nucleus collisions
does not require production of two heavy quark pairs in the same elementary collision.
Heavy quarks and antiquarks originating from different nucleon-nucleon collisions can recombine and form 
a $B_c$\ meson. If such regeneration is possible, it leads to significant enhancement of $B_c$
production in nucleus-nucleus collisions relative to proton-proton ones.

It was suggested that, due to the regeneration, the $B_c$\ meson can be observed in Au + Au collisions at RHIC, although its observation in proton-proton collisions at the same energy is hardly possible~\cite{Schroedter:2000ek}. This prediction has yet to be verified in experiments. 

The high beam energy of LHC makes it possible to measure $B_c$\ mesons in both p + p and Pb+Pb collisions.  The cross section for bottom quark production in proton-proton collisions is about one order of magnitude smaller than that for charm quarks, while the cross sections of $B_c$\ and $\Upsilon$\ are about two orders of magnitude less than that of  $J/\psi$~\cite{Averbeck:2011ga, ALICE:2011aa, Emerick:2011xu, Nigmanov:2009gu}. For a very rough estimation, the yield of a meson in the regeneration process can be assumed to be proportional to the yield of each of its constituent quarks. Therefore if regeneration is present in nucleus-nucleus collisions, the yield of $\Upsilon$\ mesons would remain two order of magnitude smaller than that of $J/\psi$, while the multiplicity of $B_c$\ mesons should be only one order of magnitude smaller in comparison to $J/\psi$. Thus the nuclear modification factor $R_{AA}$\ for $B_c$\ can be roughly one order of magnitude larger than that of $J/\psi$\ and $\Upsilon$, that is, $R_{AA}>1$, which implies enhancement of $B_c$ instead of suppression.  If such an effect is found, it will be a firm confirmation of the regeneration. 
\par
In this paper, we will calculate the yield of $B_c$\ mesons in central Pb+Pb collisions in the statistical coalescence model, and then discuss more properties including the momentum dependence in a detailed transport model. We take $\hbar=c=k_{B}=1$\ in the following.

\section{Baseline}
$B_c$\ mesons are similar to quarkonia in the sense that they are composed of heavy quarks, so that the nonrelativistic approximation can be applied to describe the interaction between them. On the other hand, the decay modes of $B_c$\ are totally different from those of quarkonia.
For excited $B_c$\ states with mass below the $B+D$\ threshold, the  only decay mode, except weak decay, is feeding down to the lower states until the ground state is reached, due to the conservation of charm and bottom. Thus the contribution from excited states is even more significant than in the case of quarkonia. The properties of ground and excited $B_c$\ mesons are discussed in different potential models~\cite{Chen:1992fq}. In order to be simple and to generalize it to finite temperature later in the transport model, we calculate the spectrum of $B_c$\ by solving the Schr\"odinger equation with a Cornell potential $V(r)=-\alpha/r+\sigma r$. By using this method, the mass spectra of charmonia and bottomonia can be well reproduced with the parameters $\alpha=\pi/12$, $\sigma=0.2$\ GeV$^2$, $m_c=1.25$\ GeV, and $m_b=4.25$\ GeV~\cite{Satz:2005hx}. With exactly the same parameters, we can obtain the mass spectrum of $B_c$\ below the threshold as $m_{B_c}(1S) = 6.36$\ GeV, $m_{B_c}(1P) = 6.72$\ GeV, $m_{B_c}(2S) = 6.90$\ GeV, and $m_{B_c}(1D) = 6.98$\ GeV. Since the binding energy of the one-dimensional state in vacuum is only about $170$\ MeV$\sim T_c$, we neglect the contribution of this state in the following for simplicity. The contribution of this state in the statistical coalescence model is only a few percent. Note that there are already experimental data for $B_c^+$ with quantum number $J^P=0^-$\ and mass $m=6.28$\ GeV~\cite{Nakamura:2010zzi}; we use the experimental value instead in the following for the $1S$ state.
\par
The cross section of $B_c^+$\ at $\sqrt{s}=1.96$\ TeV as measured by the CDF Collaboration is $d\sigma/dy\ (p_T>6 \textrm{ GeV})=15.5\pm5.0 \textrm{ nb}$\ with rapidity $|y|<1$~\cite{Nigmanov:2009gu}.
According to PYTHIA~\cite{Sjostrand:2000wi}, the cross section for $p_T>0$\ at LHC energy $\sqrt{s}=2.76$\ TeV is larger by the factor of about 4. Thus we obtain the inclusive cross section of $B_c$\ as $d\sigma/dy(p_T > 0)$ = 62 nb. The cross section of charm and bottom are estimated as $\left.d\sigma_{c\bar{c}}/dy\right|_{y=0}=620$\ $\mu$b~\cite{Averbeck:2011ga, ALICE:2011aa} and $\left.d\sigma_{b\bar{b}}/dy\right|_{y=0}=20$\ $\mu$b~\cite{Emerick:2011xu}, respectively. The elliptic flow and high $p_T$\ suppression of open heavy flavors imply strong interaction between heavy quarks and the fireball~\cite{Adare:2010de}; therefore we simply take the thermal momentum distribution for the heavy quarks.
\par
In the transport model, the momentum distribution of the initially produced $B_c$s is required. It is parametrized in the power-law form
\begin{eqnarray}
	\left.\frac{d^2\sigma}{dydp_T}\right|_{y=0}(p_T)= \frac{2(n-1)}{(n-2)\langle p_T^2 \rangle} p_T\left(1+\frac{p_T^2}{(n-2)\langle p_T^2 \rangle}\right)^{-n}\left.\frac{d\sigma}{dy}\right|_{y=0},
\end{eqnarray}
with $n=4.16$\ and $\langle p_T^2 \rangle=25.1$\ GeV$^2$\ estimated from PYTHIA~\cite{Sjostrand:2000wi} simulation. To consider the reaction rates of different $B_c$\ states, the cross section and the branch ratios of each of them are necessary. Since all the excited states feed down to the ground state without weak decays considered, we take the branch ratio as 100\%. The ratio of direct production cross section of $\psi'$\ to $J/\psi$\ is around $0.23$, while the same ratios for $\Upsilon(2S)$\ and $\Upsilon(3S)$\ to $\Upsilon(1S)$\ are both above $0.35$~\cite{Vogt:2010aa}. As a rough estimation, we take the cross section of each excited component of $B_c$\ as $0.3$\ times of the ground state; that is $\sigma^{dir}(1S):\sigma^{dir}(1P):\sigma^{dir}(2S)=10:9:3$, where we have counted the degeneracy. To include the Cronin effect, a Gauss smearing is used to modify the initial momentum dependence so that the mean $p_T^2$\ of the $J/\psi$\ in $A+A$\ collisions is larger than that in $p+p$\ collisions, that is $\langle p_T^2 \rangle_{AA}=\langle p_T^2 \rangle_{pp}+a_{gN}l$, where $l$\ is the total path length of the path that the gluons pass through the nuclei before merging into a $B_c$, with the broadening factor $a_{gN}$\ taken as $0.2\textrm{ GeV}^2\textrm{/fm}$.  Since the initially produced $B_c$s suffer strong suppression, the dependence of the final observations on the parameters discussed in this paragraph for the initial production is very weak. 

\section{Statistical Coalescence Model}
The statistical hadronization model (SHM) and statistical coalescence model (SCM) have made great success in light and heavy hadrons respectively with few parameters~\cite{BraunMunzinger:2001ip, BraunMunzinger:2000px}. 
This suggested the 
idea that hadronization of heavy hadrons can also be described within the statistical 
approach. 
In contrast to light quarks, the masses of $c$ and $b$ quarks are much larger than 
the typical temperature of the fireball. Therefore, production of $c\bar{c}$ and $b\bar{b}$
pairs at the thermal stage of the reaction can be neglected, though pre-equalibration can slightly alter the picture for $c\bar{c}$ production at LHC energy~\cite{Uphoff:2010sh}. Practically all heavy flavor pairs are produced at the initial stage of the nucleus-nucleus reaction 
in hard parton collisions. It is assumed that their numbers remains approximately unchanged 
during the fireball evolution.
\par
The basic idea of SCM for heavy flavors is expressed in the following balance equation~\cite{ BraunMunzinger:2000px}:
\begin{eqnarray}
	N_{Q\bar{Q}}^{dir}&=&\frac{1}{2}g_QN_{oQ}^{th} + g_Q^2 N_{Q\bar{Q}}^{th},
\end{eqnarray}
where $N_{Q\bar{Q}}^{dir}$\ is the number of directly produced heavy quarks, and $N^{th}_{oQ}$\ and $N^{th}_{Q\bar{Q}}$\ are numbers of open and hidden heavy flavors for the hadron gas in complete (including heavy flavors) chemical equilibrium and zero value of the corresponding heavy flavor chemical potential.  A fugacity $g_Q$\ is introduced to describe both the conservation of $Q$\ and that of $\bar{Q}$. When $N_{Q\bar{Q}}$\ is small, the event-by-event fluctuation becomes important. This effect can be included by a modification factor $(1+1/N_{Q\bar{Q}})$\ in the hidden part, that is $N_{Q\bar{Q}}=(1+1/N_{Q\bar{Q}}^{dir})g_Q^2N_{Q\bar{Q}}^{th}$, as long as the heavy flavor is mainly open heavy hadrons~\cite{Gorenstein:2000ck}. This modification leads to results similar to those of the widely used canonical ensemble modification~\cite{Rafelski:1980gk} with a deviation within several percent. In order to take the $B_c$\ meson into consideration, we generalize the above equation into the following form:
\begin{eqnarray}
	N_{c\bar{c}}^{dir}&=&\frac{1}{2}g_c(N_{oc}^{th}+g_bN_{B_c}^{th}) + \left(1+\frac{1}{N_{c\bar{c}}^{dir}}\right)g_c^2 N_{c\bar{c}}^{th},
	\label{eq_bc_1}\\
	N_{b\bar{b}}^{dir}&=&\frac{1}{2}g_b(N_{ob}^{th}+g_cN_{B_c}^{th}) + \left(1+\frac{1}{N_{b\bar{b}}^{dir}}\right)g_b^2 N_{b\bar{b}}^{th},
	\label{eq_bc_2}
\end{eqnarray}
where $N_{oc}$\ and $N_{ob}$\ are the yield of hadrons with charm and bottom except $B_c$, respectively. This formula can easily be generalized for hadrons with even more heavy quarks. The leading order approximation for $B_c$\ is simply
\begin{equation}
   N_{B_{c}} =    \frac{(2N_{c\bar{c}}^{dir})(2N_{b\bar{b}}^{dir}) }{N_{ob}^{th}N_{oc}^{th}}N_{B_{c}}^{th}.
\label{NBc}
\end{equation}
Higher orders account for the strict conservation of heavy quarks and the balance between open and hidden heavy flavors. In the following calculation, the full form Eqs.(\ref{eq_bc_1}) and (\ref{eq_bc_2}) is solved.
\par
The temperature and volume of the fireball are taken as $T=164\textrm{ MeV}$\ and $V_{\Delta y=1}=4160\textrm{ fm}^3$\ respectively~\cite{Andronic:2008gu} for central Pb+Pb collisions at $\sqrt{s}=2.76\ A$~TeV, and $N_{c\bar{c}}$\ and $N_{b\bar{b}}$\ are derived from the Glauber model with the cross sections discussed previously. All the listed charm and bottom hadrons with a mass $m$\ and spin $J$\ in the particle list from the Particle Data Group~\cite{Nakamura:2010zzi} are included. The results are as follows: $ g_c = 31.1$, $g_b = 2.39\times 10^8$, $N_{ob}/(2N_{b\bar{b}}^{dir})=98.3\%$, $N_{B_c^+}/N_{b\bar{b}}^{dir}=1.08\%$, and $N_{\Upsilon s}/N_{b\bar{b}}^{dir}=0.60\%$.\footnote{The value of the ratio $N_{\Upsilon s}/N_{b\bar{b}}^{dir}$\ predicted by SCM gives the nuclear modification factor of hidden bottom mesons $R_{AA}(\Upsilon s)=0.5\textrm{-}1$. For the ground state, $R_{AA}(\Upsilon_{1S})=1.2\textrm{-}2.5$\ is about $3\textrm{-}6$\ times larger than the experimental result \cite{CMS:2012fr}. This suggests that the hidden bottom mesons might not be completely thermalized.} Note that 
\begin{eqnarray}
	\frac{N_{B_c}^{dir}}{N_{b\bar{b}}^{dir}}&=&\frac{\sigma_{B_c}^{dir}}{\sigma_{b\bar{b}}^{dir}} =\frac{62\ \textrm{nb}}{20\ \mu \textrm{b}}=0.31\%.
\end{eqnarray}
The nuclear modification factor can be calculated as
\begin{eqnarray}
	R_{AA}&=&\frac{N_{B_c}}{N_{B_c}^{dir}}=\frac{N_{B_c}/N_{b\bar{b}}^{dir}}{N_{B_c}^{dir}/N_{b\bar{b}}^{dir}}=\frac{1.08\%}{0.31\%}=3.5.
\end{eqnarray}
Only the scalar component of $B_c$\ mesons has been observed in experiments. If we take the probably existent vector state of $B_c$\ and the isospin partner of $B^*$\ into consideration, the results are modified as follows: $ g_c = 31.1$, $g_b = 1.83\times 10^8$, $N_{ob}/(2N_{b\bar{b}}^{dir})=96.4\%$, $N_{B_c^+}/N_{b\bar{b}}^{dir}=3.30\%$, and $N_{\Upsilon s}/N_{b\bar{b}}^{dir}=0.35\%$, which results in $R_{AA}=3.30/0.31=11$. Furthermore, if $B_c(1P)$\ and $B_c(2S)$\ states are also considered, then $R_{AA}$\ can be as large as $13$.
Thus from the SCM, we do expect an enhancement of $B_c$\ mesons in central Pb+Pb collisions at LHC energy with the $R_{AA}=3.5\textrm{-}13$. 
\section{Transport Model}
In contrast to the statistical coalescence model, in which statistical equilibration of the heavy quark distribution among the hadrons at chemical freeze-out at hadronization is assumed, the transport model takes into account the deviation from the statistical equilibrium and traces the whole evolution of the fireball, and thus it can give more detail on the various processes in the QGP.  In this model, the distribution of $B_c$\ in phase space is described by the function $f_{B_c}({\bf x}, {\bf p}, t)$\ satisfying the transport equation
\begin{eqnarray}
	(\partial_t + {\bf v}\cdot{\bf {\nabla}}) f_{B_c} = - \alpha f_{B_c} + \beta,
	\label{eq_transport}
\end{eqnarray}
where ${\bf v}={\bf p}/E_{B_c}$ is the velocity of $B_c$. The dissociation rate can be expressed as
\begin{eqnarray}
\alpha({\bf x}, {\bf p}, t) &=&  
\frac{1}{E_{B_c}}\int\frac{d{\bf k}}{(2\pi)^3E_k}k_{\mu}p^{\mu}f_g^{th}\sigma\frac{\theta(T-T_c)}{\theta(T_d-T)},
\label{eq_alpha}
\end{eqnarray}
where $f_g^{th}(k, u, T)$\ is the thermal distribution of gluons and $\sigma(k, p)$\ is the cross section of the gluon dissociation process $B_c^++g\rightarrow \bar{b}+c$, which is obtained by replacing the heavy quark mass $m_Q$\ with twice the reduced mass $2\mu$\ and substituting the proper binding energy in the cross sections for quarkonia obtained by the OPE (Operator Production Expansion) method~\cite{Peskin:1979va, Bhanot:1979vb, Arleo:2001mp, Schroedter:2000ek}.  In the $B_c$\ rest frame, these can be expressed as~\cite{Peskin:1979va, Bhanot:1979vb, Arleo:2001mp, Schroedter:2000ek}
\begin{eqnarray*}
	\sigma_{1S}(\omega)&=&A_0\frac{(\omega/\epsilon_{1S}-1)^{3/2}}{(\omega/\epsilon_{1S})^5},\\
	\sigma_{1P}(\omega)&=&4A_0\frac{(\omega/\epsilon_{1P}-1)^{1/2}(9(\omega/\epsilon_{1P})^2-20(\omega/\epsilon_{1P})+12)}{(\omega/\epsilon_{1P})^7},\\
	\sigma_{2S}(\omega)&=&16A_0\frac{(\omega/\epsilon_{2S}-1)^{3/2}(\omega/\epsilon_{2S}-3)^2}{(\omega/\epsilon_{2S})^7},
	\label{eq_cross_section}
\end{eqnarray*}
where $\omega$ is the gluon energy, $A_0=2^{11}\pi3^{-3}(2\mu)^{-3/2}\epsilon_{1S}^{-1/2}$, $\mu=(m_bm_c)/(m_b+m_c)$\ is the reduced mass, and $\epsilon_{J}$ is the binding energy of the $B_c$\ state $J$.  Here we use vacuum values; that is, the binding energy is found as a difference between
the sum of the masses of the ground state $B$\ and $D$\ mesons and the mass of the corresponding $B_c$\ state.\footnote{In reality, the binding energies of the $B_c$\ state in a deconfined medium are modified due to the screening effect. However, in a recent study \cite{Emerick:2011xu}, the difference in Upsilon suppression between different binding energies---strong binding as in our calculation and weak binding in which heavy quarks are totally thermalized---is discussed and compared; it is found that the strong-binding calculation better explains the experimental data from RHIC and LHC. Based on these considerations, we take the binding energy in the cross section from the value in vacuum as an approximation.} To take into account the recoil effect due to the finite mass of $B_c$, we further replace the binding energy $\epsilon$\ by the threshold $\omega_0=\left[1+\epsilon/(2m_{B_c})\right]\epsilon$\ as in~\cite{Polleri:2003kn}.  
\par
\begin{figure}[!htb]
	\centering
	\includegraphics[width=0.8\textwidth]{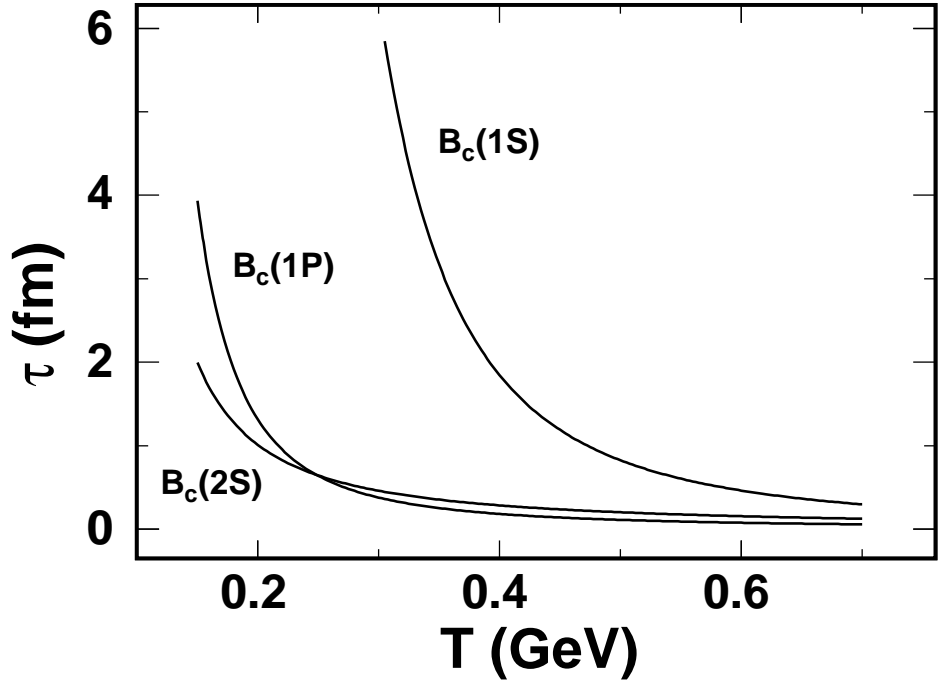}
	\caption{Lifetime of $B_c$\ meson  with account for the gluon dissociation in QGP. 
The velocity of the QGP is taken as $v=0$\ in this figure.}
	\label{fig_lifetime}
\end{figure}

 
The lifetime of $B_c$\ with account for the gluon dissociation is shown in Fig.~\ref{fig_lifetime}.  The binding energy of the ground state is much larger than that of the excited states, and therefore the ground state lives longer. Comparing the lifetime of the $B_c$\ meson to the radius of the nucleus $R(\textrm{Pb})\sim 6$\ fm, one finds that the ground state suffers strong suppression at a temperature above $400$\ MeV and little suppression below $300$\ MeV, while the suppression to the excited states is already very strong at $200$\ MeV.  
\par
The theta functions in the numerator and denominator in (\ref{eq_alpha}) restrict the above process between the critical temperature $T_c$\ of light hadrons and the dissociation temperature $T_d$\ of $B_c$\ mesons. Below $T_c$, there are no gluons in the fireball.  Above $T_d$, the bottom and charm quarks are strongly screened, and can no longer form a bound state.
\par
With the heavy quark potential at finite temperature obtained from lattice QCD~\cite{Kaczmarek:2005ui}, the dissociation temperature $T_d$\ of $B_c$\ mesons can be calculated by solving the Schr\"odinger equation as in~\cite{Satz:2005hx}. The results are listed in Table ~\ref{tb_td}.  The dissociation temperature is calculated for two extreme cases.  In the first one, the potential is assumed to be equal to the internal energy: $V=U$; in the other one it is equal to the  free energy: $V=F$. Note that the difference between them is the entropy term.  The former corresponds to a quick adiabatic screening, while in the latter a strong heat exchange between the heavy quark system and the medium is assumed~\cite{Shuryak:2004tx}.  Since the difference is significant, we will do calculations in both limits in parallel.
\begin{table}[!hbt]
	\centering
	\begin{tabular}{|c|c|c|c|}
		\hline
		States of $B_c$		& $1S$ & $1P$ & $2S$ \\
		\hline
		$T_d$/$T_c$ ($V=U$)	& 3.27 & 1.59 & 1.41\\
		\hline
		$T_d$/$T_c$ ($V=F$)	& 1.51 & -    & -\\
		\hline
	\end{tabular}
\caption{The dissociation temperature $T_d$\ of $B_c$\ mesons scaled by the critical temperature 
$T_c$. The heavy quark potential $V$\ is considered as the internal energy 
$U$\ and the free energy $F$\ at finite temperature, respectively. 
The dash means the meson does not survive above $T_c$.}
	\label{tb_td}
\end{table}
\par
Besides the dissociation of $B_c$, the inverse process $\bar{b}+c\rightarrow B_c+g$\ is also considered in (\ref{eq_transport}), which lies in the regeneration rate
\begin{displaymath}
   \beta=\frac{1}{2E_{B_c}}\int\frac{d\vec{k}}{(2\pi)^32E_g}\frac{d\vec{q}_c}{(2\pi)^32E_c}\frac{d\vec{q}_{\bar b}}{(2\pi)^32E_{\bar b}}W(s)f_cf_{\bar{b}}(1+f_g)(2\pi)^4\delta^{(4)}(p+k-q_c-q_{\bar b}),
\end{displaymath}
where $p$, $k$, $q_c$, and $q_{\bar b}$ are the momenta of the $B_c$, the gluon, the charm quark and the bottom quark, respectively. $W(s)$\ is the transition probability, which is related to that of the dissociation process by detailed balance of the cross section entering in Eq. (\ref{eq_alpha}). $f_c$, $f_{\bar b}$, and $f_g$\ are distribution functions of $c$\ and $b$\ quarks and gluons. $f_g$\ is taken as a thermal distribution as in the dissociation rate $\alpha$. For simplicity, we also assume kinetic thermalization for the heavy quarks, and we neglect the Pauli blocking effect. That is the distribution function $f_Q=\rho_Qf_Q^{th}$, where the momentum part $f_Q^{th}$\ is the normalized Boltzmann distribution, and the density of heavy quarks, $\rho_Q$\ is determined by the conservation law of heavy quarks
\begin{eqnarray}
   \partial_{t}\rho_Q+\nabla\cdot(\rho_Q \vec{v})&=&0,
\end{eqnarray}
where $\vec{v}$\ is taken as the velocity of the medium. The initial condition is determined by the Glauber model. The regeneration process takes place in the temperature window $T_c<T<T_d$. When $T<T_c$, there are no partons, and therefore the previous process does not exist. When $T>T_d$, the dissociation rate $\alpha$\ is infinity, and the $B_c$ gets dissociated as soon as it forms; therefore  regeneration does not occur at high temperature, either. Since the dissociation temperature $T_d$\ depends strongly on the form of the heavy quark potential, one would expect that the yield of $B_c$\ from the regeneration process also depends strongly on the potential.
\par
As a background of the $B_c$\ suppression and regeneration, the fireball is described by (2+1)-dimensional ideal hydrodynamics with the assumption of boost invariance~\cite{Kolb:2000fha}.  The equations of state (EoS) are taken as a massive ideal gas of hadrons and partons in confined and deconfined phases, respectively, with a first-order phase transition at $T_c=165$\ MeV as was done for $J/\psi$\ and $\Upsilon$~\cite{Zhu:2004nw}.
Multiplying this critical temperature by the ratio $T_d/T_c$\ in Table \ref{tb_td}, one obtains the dissociation temperature $T_d$, regardless of the absolute value  of $T_c$\ in the lattice simulation.  Otherwise, there could be even more enhancement of $B_c$. The initial condition is decided by the Glauber model, and the maximum temperature of the fireball in the most central collision is $485$\ MeV at a thermalization time $\tau_0=0.6$\ fm.
\par
\begin{figure}[!htb]
	\centering
	\includegraphics[width=0.8\textwidth]{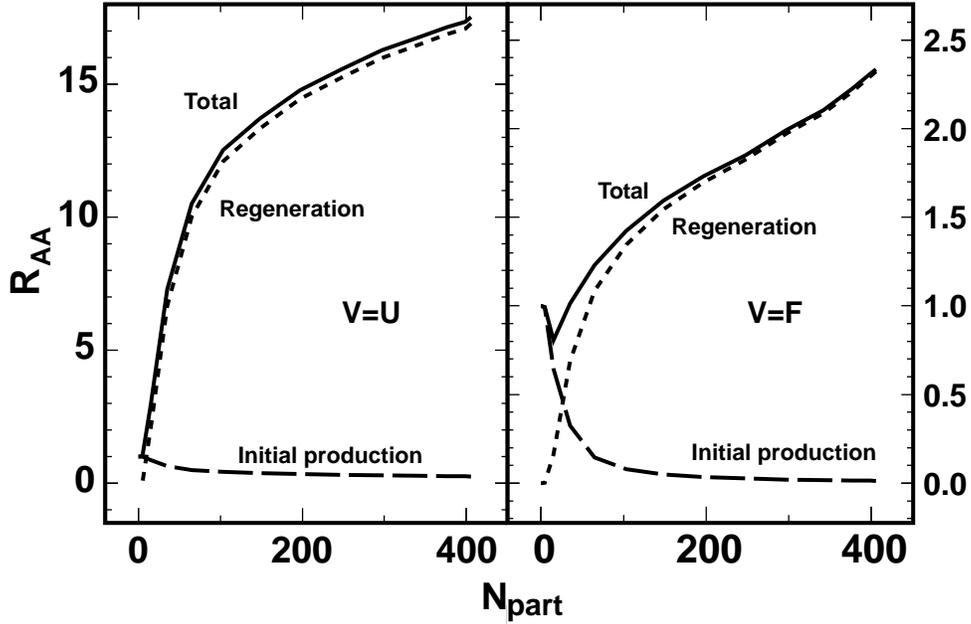}
	\caption{The nuclear modification factor $R_{AA}$\ of $B_c$\ meson as a 
function of the participant nucleon number $N_{\textrm{part}}$\ in $V=U$(a) and 
$V=F$(b) limits.  The long-dashed, dashed, and solid curves are the contribution from initial production, regeneration and the total respectively.}
	\label{fig_raa_np}
\end{figure}

The centrality dependence of the  nuclear modification factor $R_{AA}$\ in both limits of $V=U$\ and $V=F$\ are shown in Fig.~\ref{fig_raa_np}.  If regeneration is ignored, $R_{AA}$\ of the initially produced $B_c$\ is $0.25$\ for $V=U$\ and $0.01$\ for $V=F$, which are reasonable when compared to the experimental results for $\Upsilon$.  The $R_{AA}$\ values of $\Upsilon(1S+2S+3S)$\ at RHIC~\cite{Reed:2011fr} and $\Upsilon(1S)$\ at LHC~\cite{Chatrchyan:2012np} in central collisions are consistent with the simple assumption that the ground state survives while all the excited states melt in the fireball. One can expect that the suppression of $B_c$\ is stronger than that of $\Upsilon$. In the $V=U$\ limit, the maximum temperature is still lower than $T_d$\ of $B_c(1S)$, and about half of the $B_c(1S)$\ mesons survive the gluon dissociation, which is less than that of $\Upsilon(1S)$. In the $V=F$\ limit, the dissociation temperature is much lower, and the ground state suffers even stronger suppression. 
\par
In both cases, the contribution from the initial production is two orders of magnitude smaller than that from the regeneration, and thus the population of $B_c$\ is dominated by the latter. Note that a larger $R_{AA}$\ value for $B_c$ than that for $\Upsilon$\ already implies the regeneration mechanism in the fireball. In both limits of our calculations, the $R_{AA}=2.3\textrm{-}17.5$, exceeding unity. The temperature for regeneration in the transport model is higher than that in SCM; thus one cannot ascertain whether the yield of $B_c$\ is smaller in the transport model or in the SCM. Actually, this interval covers the SCM results and also implies strong enhancement.
The remarkable difference between the two limits comes from two factors. First, the excited states does not survive at the free energy limit, while they play an important role in the internal energy limit. In the most central collisions, the contribution from the ground state and the exited states are almost the same for $V=U$. Second, the dissociation temperature of the ground state is much higher in the internal energy limit, which 
allows more regeneration in the hot fireball. 
\par
When the regeneration inside the fireball is confirmed, the nuclear modification factor loses its meaning as a survival probability, because the observed $B_c$\ mesons are mostly regenerated instead of surviving. Thus $R_{AA}$\ depends not only on how the $B_c$\ mesons interact with the medium but also on the production cross section of $B_c$\ and heavy quarks in hard nucleon-nucleon collisions.
\par
To characterize the producing and/or surviving ability of $B_c$\ from the regeneration, we take a ratio of the final yield of $B_c$\ mesons over those of the heavy quarks in the unit rapidity region~\cite{Liu:2009wza} as
\begin{eqnarray}
	F&\equiv&\frac{dN_{B_c}/dy}{dN_{c}/dy\cdot dN_{b}/dy}.
	\label{eq_s}
\end{eqnarray}
While $R_{AA}$\ compares the final yield of $B_c$\ to the initial yield, $F$\ compares it to the source of regeneration, and thus it  becomes independent of the initial cross sections of $B_c$\ and heavy quarks when the regeneration dominates. If there is no nuclear matter effect, the fraction is just
\begin{eqnarray}
   f_{\textrm{Glauber}}&=& \frac{d\sigma_{B_c}/dy\cdot \sigma_{in}\ \ \ \ \ }{d\sigma_{c}/dy\cdot d\sigma_{b}/dy}\frac{1}{N_{\textrm{coll}}},
   \label{eq_f_glauber}
\end{eqnarray}
where $\sigma_{B_c}$, $\sigma_c$, $\sigma_b$, and $\sigma_{in}$\ are the cross sections of $B_c$, $c$\ and $b$\ quarks, and the inelastic collisions in $pp$ collisions, respectively, and $N_{\textrm{coll}}$\ is the number of binary collisions. Under our assumption that the heavy quarks are conserved in the fireball, the denominators in $F$\ and $F_{\textrm{Glauber}}$\ are the same in (\ref{eq_s}), and thus $R_{AA}=F/F_{\textrm{Glauber}}$. The comparison of $F$\ between different beam energies for regeneration-dominant processes makes more sense than $R_{AA}$, since the initial cross sections in $F_{\textrm{Glauber}}$\ is dropped.

\begin{figure}[!htb]
	\centering
	\includegraphics[width=0.8\textwidth]{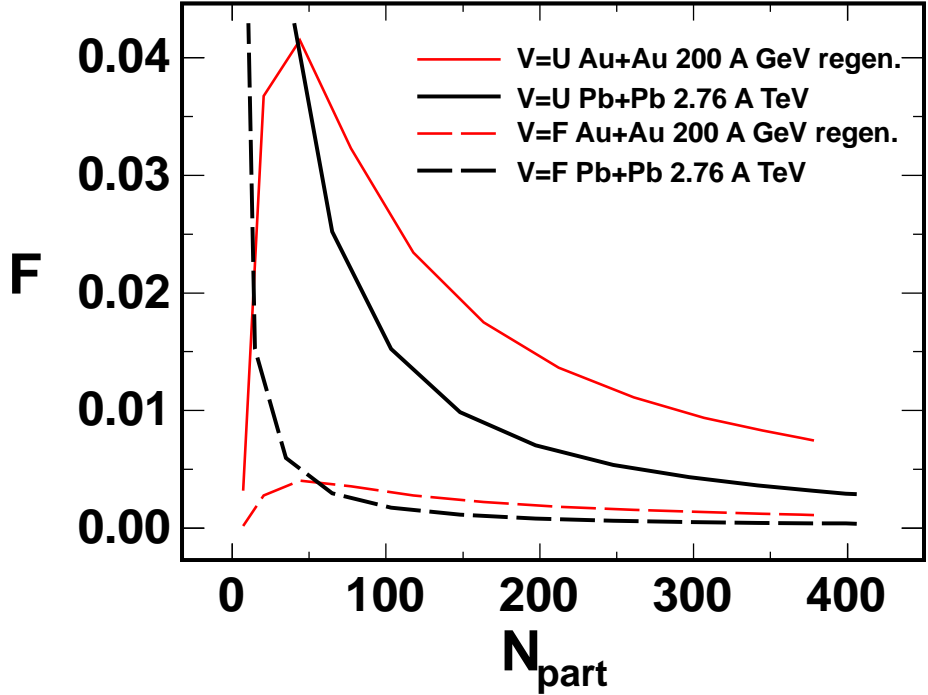}
	\caption{(Color online) The final ratio $F$(defined in (\ref{eq_s})) of $B_c$\ meson as a function of the participant nucleon number $N_{\textrm{part}}$. The long-dashed and solid lines are $V=F$\ and $V=U$\ limit respectively. The black thick lines are for LHC energy, while the red thin lines are for the regenerated $B_c$\ mesons at RHIC energy as a comparison.}
	\label{fig_s_np}
\end{figure}

The final ratio $F$\ at LHC energy is shown in Fig.~\ref{fig_s_np} compared with that of the regenerated $B_c$\ at RHIC. The final ratio is larger in the $V=U$\ case, as one would expect from the dissociation temperatures. Since the temperature at LHC is much higher, the final ratio is smaller than the corresponding case at RHIC, which is consistent with the picture of color screening. The smaller $F$\ in central collisions also implies smaller producing and/or surviving ability of $B_c$\ in a hot, large medium, and the growth of $R_{AA}$\ with $N_{\textrm{part}}$\ is mainly due to the increasing multiplicity of heavy quarks. Meanwhile the yield is very sensitive to the potential as we found in $R_{AA}$, so that $F$\ with $V=F$\ at RHIC is even smaller than that with $V=U$\ at LHC. The difference in behavior between RHIC and LHC at small $N_{\textrm{part}}$\ is artificial. That is, because the initial production is not included in calculations for RHIC in Fig.~\ref{fig_s_np}, $F$\ as a measurement of producing and/or surviving ability only makes sense when the regeneration dominates.
\par
\begin{figure}[!htb]
   \begin{center}
	\includegraphics[width=0.8\textwidth]{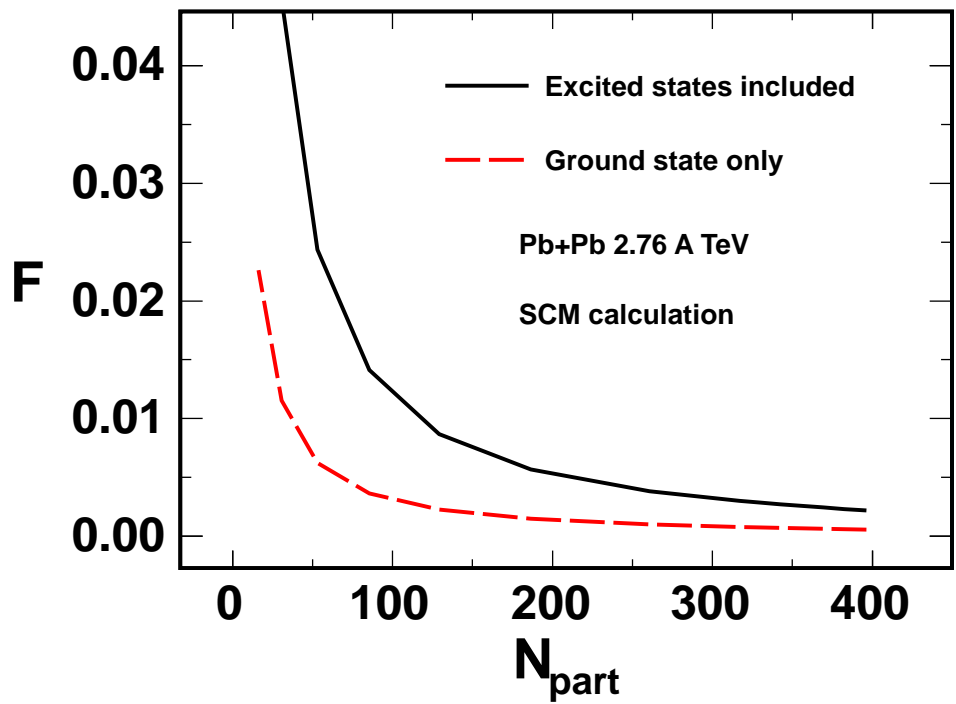}
   \end{center}
   \caption{(Color online) The final ratio $F$\ [defined in (\ref{eq_s})] of $B_c$\ meson as a function of the participant nucleon number $N_{\textrm{part}}$\ from SCM calculation. The red long-dashed and black solid lines are calculations of observed scalar $B_c$\ state only and that including excited $B_c$\ states (as listed in Table~\ref{tb_td}), respectively.}
   \label{fig_s_np_scm}
\end{figure}

The centrality dependence of $F$\ in SCM is similar, as shown in Fig.~\ref{fig_s_np_scm}. In order to apply the SCM model to different centralities, we assumed that the volume of the fireball is proportional to the multiplicity of charged particles, which is measured by experiments~\cite{Aamodt:2010cz}. The yield of heavy quarks is from a Glauber model as before. The result can be understood from (\ref{NBc}), which implies in unit rapidity
\begin{displaymath}
   F=\frac{4N_{B_c}^{th}}{N_{oc}^{th}N_{ob}^{th}}=\frac{4n_{B_c}^{th}}{n_{oc}^{th}n_{ob}^{th}}\frac{1}{V}\propto\frac{1}{V}.
   \label{eq_s0}
\end{displaymath}
The volume increases with $N_{\textrm{part}}$, and thus the final ratio $F$\ decreases, which implies that it is more difficult for a given pair of heavy quarks to meet and combine with each other. When considering $F=V$, there is almost $N_p$\ scaling, with a deviation within a few percent in peripheral collisions. The deviation comes from the change in heavy quark density and the event-by-event fluctuation. In the most central collisions, F is between $6\times 10^{-4}$\ and $2\times 10^{-3}$, which is consistent with the results ($7\times 10^{-5})\textrm{-}(3\times 10^{-3})$\ in previous literature~\cite{Kuznetsova:2006bh} at volume $dV/dy=600\textrm{-}800$\ fm and temperature $T=140\textrm{-}260$\ MeV.
\par
\begin{figure}[!htb]
	\centering
	\includegraphics[width=0.8\textwidth]{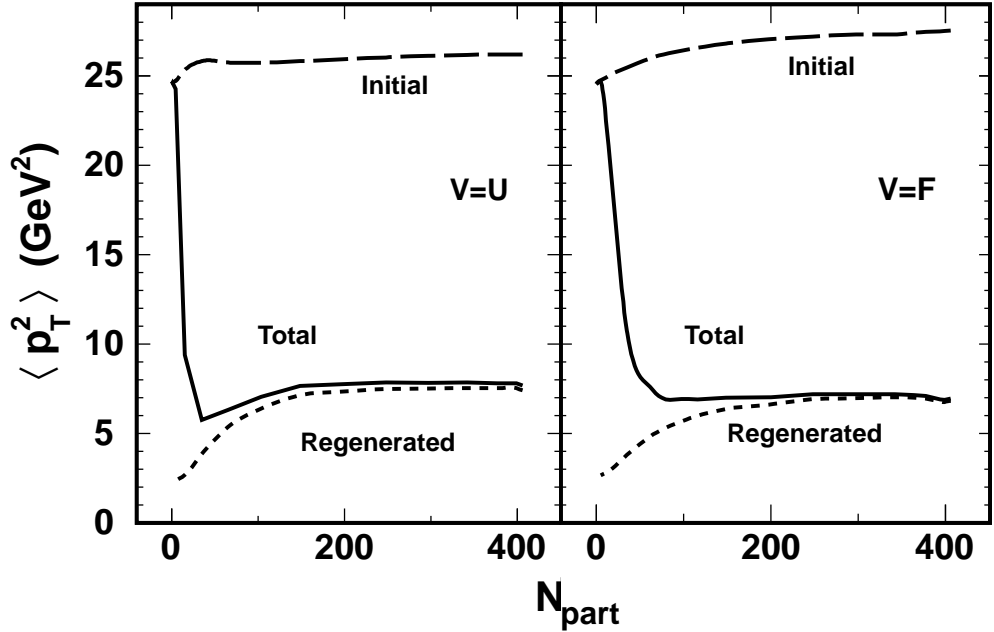}
	\caption{The average transverse momentum square $\langle p_T^2 \rangle$\ as a function of the participant nucleon number $N_{\textrm{part}}$\ in $V=U$(a) and $V=F$(b) limits. The long-dashed, dashed, and solid curves are for initial production, regeneration and the total, respectively.}
	\label{fig_pt2}
\end{figure}

One of the advantages of the transport model is that it allows one to investigate the momentum of particles, which is sensitive to the dynamics. The initially produced $B_c$\ mesons come from the hard collisions, while the regenerated $B_c$s are merged from the heavy quarks that are softened by the medium. The typical energy scale of the medium is the temperature, which is much smaller than the typical energies of hard protons in the initial beams. Thus the transverse momentum carried by the initial production is obviously larger than that of the regeneration. When the regeneration becomes important, there is a suppression in $\langle p_T^2 \rangle$, as shown in Fig.~\ref{fig_pt2}. The same phenomenon is also expected for $J/\psi$\ at LHC~\cite{Liu:2009nb}.
\par
\begin{figure}[!htb]
	\centering
	\includegraphics[width=0.8\textwidth]{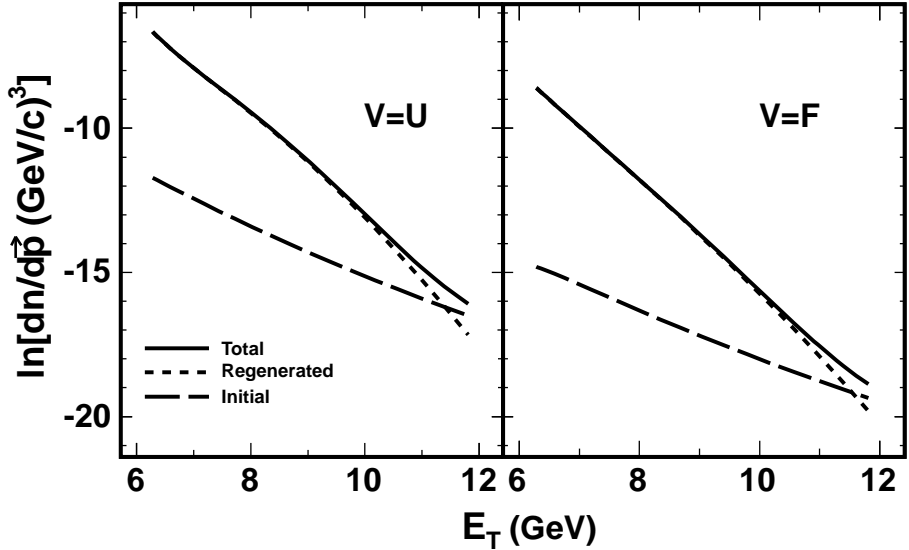}
	\caption{$B_c$\ meson spectrum with respect to the transverse energy $E_T=\sqrt{p_T^2+m^2}$. In $V=U$(a) and $V=F$(b) limits.  The long-dashed, dashed, and solid curves are for initial production, regeneration and the total, respectively.}
	\label{fig_pt_spectrum}
\end{figure}

Since thermalized heavy particles from a pointlike thermal source follow a Boltzmann distribution, that is, ${dN}/d{\bf p}\propto \exp(-E_T/T)$\ at mid-rapidity, we plot $\ln (dN/d{\bf p})$\ at mid-rapidity as a function of $E_T$\ in Fig.~\ref{fig_pt_spectrum} for central Pb+Pb collisions. The plot would be a straight line for a Boltzmann distribution. The softening of the spectrum in Pb+Pb collisions is obvious through the change of the slope. If a superposition of Boltzmann distributions at different temperatures is considered, only a concave curve is expected. However, in the internal energy limit, it is a convex curve at $E_T\sim9$\ GeV. Such a behavior is mainly attributed to the suppression of excited states. If we switch off the gluon dissociation process of the excited states, the obvious bend at low $p_T$\ disappears. The effective temperature extracted from the spectrum at low $p_T$\ according to the Boltzmann distribution is about $550$\ MeV, which is above the maximum temperature input of the fireball. This mainly results from the radial flow. When the velocity of the fireball is switched off in the regeneration process, the effective temperatures are around $250$\ MeV, while a calculation of average temperature directly from the transport model is about $230$\ MeV. Since the radial flow seems large, one would also expect large elliptic flow.
\par
\begin{figure}[!htb]
	\centering
	\includegraphics[width=0.8\textwidth]{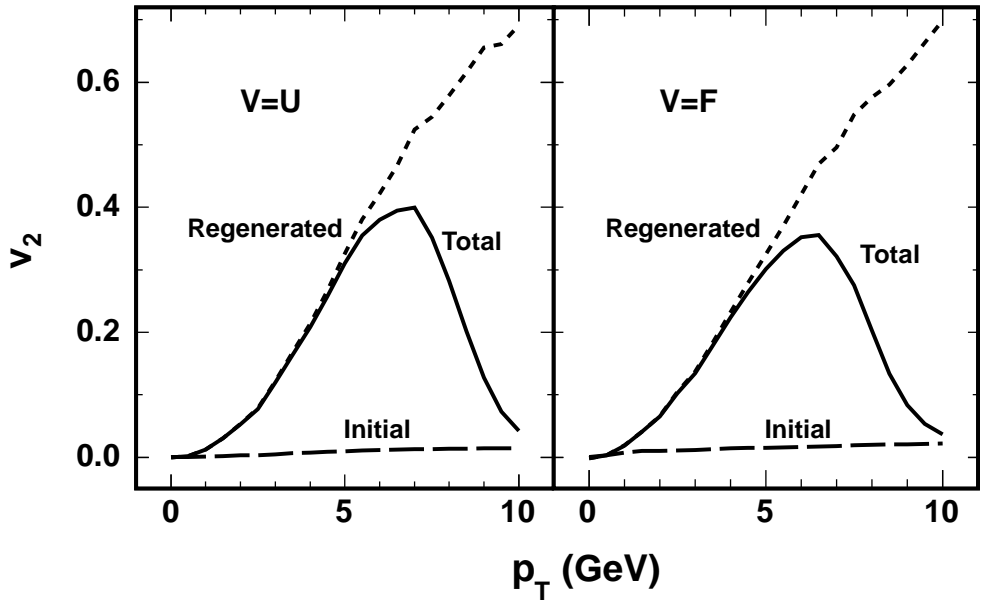}
	\caption{The elliptic flow $v_2$\ of $B_c$\ mesons as a function of the transverse momentum $p_T$\ at impact parameter $b=8.4$\ fm in Pb+Pb collisions. Two limits of the heavy quark potential $V=U$(a) and $V=F$(b) are shown. The long-dashed, dashed, and solid curves are for initial production, regenerated and total contributions, respectively.}
	\label{fig_v2}
\end{figure}

The elliptic flow at $b=8.4$\ fm is shown in Fig.~\ref{fig_v2}. The initially produced $B_c$s do not thermalize with the medium, and thus they carry a small elliptic flow. The nonzero flow comes from the suppression process similar to that for jets. The regenerated $B_c$s are born inside the fireball, and thus they inherit a relatively large flow of the medium through heavy quarks. The total flow is dominated by the regeneration at low $p_T$, and it decreases at high $p_T$\ when the initial production becomes important. Since ideal hydrodynamics and the kinetic thermalization of the heavy quarks are assumed, the flow of regenerated $B_c$\ at high $p_T$\ is not reliable. Note that the assumption of the thermalization of the momenta of bottom quarks is essential in this $v_2$\ calculation. If the bottom quarks are not thermalized, the flow of $B_c$\ can be lower than our results in the whole range of $p_T$, which is to be measured in future experiments.
\section{Conclusion}
Based on the results of both the statistical coalescence model and the transport model, an enhancement of the $B_c$\ meson yield in Pb+Pb collisions relative to $p+p$\ collisions at $\sqrt{s}=2.76$\ TeV is predicted. If such an enhancement is observed, we can conclude firmly that the regeneration of $B_c$\ mesons in the fireball occurs. Thus the $J/\psi$\ production in Pb+Pb collisions at LHC is most likely dominated by the regeneration mechanism, and the final fraction $F$\ in (\ref{eq_s0}) can be used to better characterize the hot nuclear matter effect on quarkonia in the fireball. In the opposite case, i.e., if no $B_c$\ enhancement is observed, one has to conclude that the observed $\Upsilon$\ yield is dominated by the initial production. According to our transport results, the transverse momentum of $B_c$\ is suppressed accompanying the enhancement in yield, since the regenerated $B_c$s are produced at a lower energy scale than that of  the initial production. The $p_T$\ spectrum is much softer than that in $pp$\ collisions, and in the $V=U$ limit, there is a suppression at low $p_T$\ compared to the thermal distribution, which results from the suppression of the excited states. Hopefully, the study of $B_c$\ mesons will shed more light on the production and suppression of heavy quarkonia and properties of hot quark matter.


\begin{acknowledgments}
Y.L. is grateful to Jan Uphoff and Kai Zhou for helpful discussions.
This work is supported by the Helmholtz International Center for FAIR
within the framework of the LOEWE program launched by the State of Hesse.
\end{acknowledgments}

\bibliography{ref}

\begin{thebibliography}{10}%
\makeatletter
\providecommand \@ifxundefined [1]{%
 \ifx #1\undefined \expandafter \@firstoftwo
 \else \expandafter \@secondoftwo
\fi
}%
\providecommand \@ifnum [1]{%
 \ifnum #1\expandafter \@firstoftwo
 \else \expandafter \@secondoftwo
\fi
}%
\providecommand \enquote [1]{``#1''}%
\providecommand \bibnamefont  [1]{#1}%
\providecommand \bibfnamefont [1]{#1}%
\providecommand \citenamefont [1]{#1}%
\providecommand\href[0]{\@sanitize\@href}%
\providecommand\@href[1]{\endgroup\@@startlink{#1}\endgroup\@@href}%
\providecommand\@@href[1]{#1\@@endlink}%
\providecommand \@sanitize [0]{\begingroup\catcode`\&12\catcode`\#12\relax}%
\@ifxundefined \pdfoutput {\@firstoftwo}{%
 \@ifnum{\z@=\pdfoutput}{\@firstoftwo}{\@secondoftwo}%
}{%
 \providecommand\@@startlink[1]{\leavevmode\special{html:<a href="#1">}}%
 \providecommand\@@endlink[0]{\special{html:</a>}}%
}{%
 \providecommand\@@startlink[1]{%
  \leavevmode
  \pdfstartlink
   attr{/Border[0 0 1 ]/H/I/C[0 1 1]}%
   user{/Subtype/Link/A<</Type/Action/S/URI/URI(#1)>>}%
  \relax
 }%
 \providecommand\@@endlink[0]{\pdfendlink}%
}%
\providecommand \url  [0]{\begingroup\@sanitize \@url }%
\providecommand \@url [1]{\endgroup\@href {#1}{\urlprefix}}%
\providecommand \urlprefix [0]{URL }%
\providecommand \Eprint[0]{\href }%
\@ifxundefined \urlstyle {%
  \providecommand \doi [1]{doi:\discretionary{}{}{}#1}%
}{%
  \providecommand \doi [0]{doi:\discretionary{}{}{}\begingroup
  \urlstyle{rm}\Url }%
}%
\providecommand \doibase [0]{http://dx.doi.org/}%
\providecommand \Doi[1]{\href{\doibase#1}}%
\providecommand \bibAnnote [3]{%
  \BibitemShut{#1}%
  \begin{quotation}\noindent
    \textsc{Key:}\ #2\\\textsc{Annotation:}\ #3%
  \end{quotation}%
}%
\providecommand \bibAnnoteFile [2]{%
  \IfFileExists{#2}{\bibAnnote {#1} {#2} {\input{#2}}}{}%
}%
\providecommand \typeout [0]{\immediate \write \m@ne }%
\providecommand \selectlanguage [0]{\@gobble}%
\providecommand \bibinfo [0]{\@secondoftwo}%
\providecommand \bibfield [0]{\@secondoftwo}%
\providecommand \translation [1]{[#1]}%
\providecommand \BibitemOpen[0]{}%
\providecommand \bibitemStop [0]{}%
\providecommand \bibitemNoStop [0]{.\EOS\space}%
\providecommand \EOS [0]{\spacefactor3000\relax}%
\providecommand \BibitemShut [1]{\csname bibitem#1\endcsname}%
\bibitem{Adcox:2001jp}%
  \BibitemOpen
  \bibfield{author}{%
  \bibinfo {author} {\bibfnamefont{K.}~\bibnamefont{Adcox}} \emph{et~al.}
  (\bibinfo {collaboration} {PHENIX}),\ }%
  \bibfield{journal}{%
  \Doi{10.1103/PhysRevLett.88.022301}{\bibinfo {journal} {Phys. Rev. Lett.}}\
  }%
  \textbf{\bibinfo {volume} {88}},\ \bibinfo {pages} {022301} (\bibinfo {year}
  {2002}),\
  \Eprint{http://arxiv.org/abs/nucl-ex/0109003}{arXiv:nucl-ex/0109003}%
  \bibAnnoteFile{NoStop}{Adcox:2001jp}%
\bibitem{Adare:2006ti}%
  \BibitemOpen
  \bibfield{author}{%
  \bibinfo {author} {\bibfnamefont{A.}~\bibnamefont{Adare}} \emph{et~al.}
  (\bibinfo {collaboration} {PHENIX}),\ }%
  \bibfield{journal}{%
  \Doi{10.1103/PhysRevLett.98.162301}{\bibinfo {journal} {Phys. Rev. Lett.}}\
  }%
  \textbf{\bibinfo {volume} {98}},\ \bibinfo {pages} {162301} (\bibinfo {year}
  {2007}),\
  \Eprint{http://arxiv.org/abs/nucl-ex/0608033}{arXiv:nucl-ex/0608033}%
  \bibAnnoteFile{NoStop}{Adare:2006ti}%
\bibitem{Matsui:1986dk}%
  \BibitemOpen
  \bibfield{author}{%
  \bibinfo {author} {\bibfnamefont{T.}~\bibnamefont{Matsui}}\ and\ \bibinfo
  {author} {\bibfnamefont{H.}~\bibnamefont{Satz}},\ }%
  \bibfield{journal}{%
  \Doi{10.1016/0370-2693(86)91404-8}{\bibinfo {journal} {Phys. Lett.}}\ }%
  \textbf{\bibinfo {volume} {B178}},\ \bibinfo {pages} {416} (\bibinfo {year}
  {1986})%
  \bibAnnoteFile{NoStop}{Matsui:1986dk}%
\bibitem{Gonin:1996wn}%
  \BibitemOpen
  \bibfield{author}{%
  \bibinfo {author} {\bibfnamefont{M.}~\bibnamefont{Gonin}} \emph{et~al.}
  (\bibinfo {collaboration} {NA50}),\ }%
  \bibfield{journal}{%
  \Doi{10.1016/S0375-9474(96)00373-9}{\bibinfo {journal} {Nucl. Phys.}}\ }%
  \textbf{\bibinfo {volume} {A610}},\ \bibinfo {pages} {404c} (\bibinfo {year}
  {1996})%
  \bibAnnoteFile{NoStop}{Gonin:1996wn}%
\bibitem{Spieles:1999kp}%
  \BibitemOpen
  \bibfield{author}{%
  \bibinfo {author} {\bibfnamefont{C.}~\bibnamefont{Spieles}}, \bibinfo
  {author} {\bibfnamefont{R.}~\bibnamefont{Vogt}}, \bibinfo {author}
  {\bibfnamefont{L.}~\bibnamefont{Gerland}}, \bibinfo {author}
  {\bibfnamefont{S.}~\bibnamefont{Bass}}, \bibinfo {author}
  {\bibfnamefont{M.}~\bibnamefont{Bleicher}}, \emph{et~al.},\ }%
  \bibfield{journal}{%
  \Doi{10.1103/PhysRevC.60.054901}{\bibinfo {journal} {Phys.Rev.}}\ }%
  \textbf{\bibinfo {volume} {C60}},\ \bibinfo {pages} {054901} (\bibinfo {year}
  {1999}),\ \Eprint{http://arxiv.org/abs/hep-ph/9902337}{arXiv:hep-ph/9902337
  [hep-ph]}%
  \bibAnnoteFile{NoStop}{Spieles:1999kp}%
\bibitem{Geiss:1998ma}%
  \BibitemOpen
  \bibfield{author}{%
  \bibinfo {author} {\bibfnamefont{J.}~\bibnamefont{Geiss}}, \bibinfo {author}
  {\bibfnamefont{C.}~\bibnamefont{Greiner}}, \bibinfo {author}
  {\bibfnamefont{E.}~\bibnamefont{Bratkovskaya}}, \bibinfo {author}
  {\bibfnamefont{W.}~\bibnamefont{Cassing}},\ and\ \bibinfo {author}
  {\bibfnamefont{U.}~\bibnamefont{Mosel}},\ }%
  \bibfield{journal}{%
  \Doi{10.1016/S0370-2693(98)01577-9}{\bibinfo {journal} {Phys.Lett.}}\ }%
  \textbf{\bibinfo {volume} {B447}},\ \bibinfo {pages} {31} (\bibinfo {year}
  {1999}),\ \Eprint{http://arxiv.org/abs/nucl-th/9803008}{arXiv:nucl-th/9803008
  [nucl-th]}%
  \bibAnnoteFile{NoStop}{Geiss:1998ma}%
\bibitem{Armesto:1998rc}%
  \BibitemOpen
  \bibfield{author}{%
  \bibinfo {author} {\bibfnamefont{N.}~\bibnamefont{Armesto}}, \bibinfo
  {author} {\bibfnamefont{A.}~\bibnamefont{Capella}},\ and\ \bibinfo {author}
  {\bibfnamefont{E.}~\bibnamefont{Ferreiro}},\ }%
  \bibfield{journal}{%
  \Doi{10.1103/PhysRevC.59.395}{\bibinfo {journal} {Phys.Rev.}}\ }%
  \textbf{\bibinfo {volume} {C59}},\ \bibinfo {pages} {395} (\bibinfo {year}
  {1999}),\ \Eprint{http://arxiv.org/abs/hep-ph/9807258}{arXiv:hep-ph/9807258
  [hep-ph]}%
  \bibAnnoteFile{NoStop}{Armesto:1998rc}%
\bibitem{Kahana:1999qp}%
  \BibitemOpen
  \bibfield{author}{%
  \bibinfo {author} {\bibfnamefont{D.}~\bibnamefont{Kahana}}\ and\ \bibinfo
  {author} {\bibfnamefont{S.}~\bibnamefont{Kahana}},\ }%
  \bibfield{journal}{%
  \Doi{10.1016/S0146-6410(99)00082-4}{\bibinfo {journal}
  {Prog.Part.Nucl.Phys.}}\ }%
  \textbf{\bibinfo {volume} {42}},\ \bibinfo {pages} {269} (\bibinfo {year}
  {1999})%
  \bibAnnoteFile{NoStop}{Kahana:1999qp}%
\bibitem{Qiu:2001at}%
  \BibitemOpen
  \bibfield{author}{%
  \bibinfo {author} {\bibfnamefont{J.-w.}\ \bibnamefont{Qiu}}, \bibinfo
  {author} {\bibfnamefont{J.~P.}\ \bibnamefont{Vary}},\ and\ \bibinfo {author}
  {\bibfnamefont{X.-f.}\ \bibnamefont{Zhang}},\ }%
  \bibfield{journal}{%
  \Doi{10.1016/S0375-9474(01)01430-0}{\bibinfo {journal} {Nucl.Phys.}}\ }%
  \textbf{\bibinfo {volume} {A698}},\ \bibinfo {pages} {571} (\bibinfo {year}
  {2002}),\ \Eprint{http://arxiv.org/abs/nucl-th/0106040}{arXiv:nucl-th/0106040
  [nucl-th]}%
  \bibAnnoteFile{NoStop}{Qiu:2001at}%
\bibitem{Kabana:2000iu}%
  \BibitemOpen
  \bibfield{author}{%
  \bibinfo {author} {\bibfnamefont{S.}~\bibnamefont{Kabana}},\ }%
  \bibfield{journal}{%
  \Doi{10.1088/1367-2630/3/1/316}{\bibinfo {journal} {New J.Phys.}}\ }%
  \textbf{\bibinfo {volume} {3}},\ \bibinfo {pages} {16} (\bibinfo {year}
  {2001}),\ \Eprint{http://arxiv.org/abs/hep-ph/0004138}{arXiv:hep-ph/0004138
  [hep-ph]}%
  \bibAnnoteFile{NoStop}{Kabana:2000iu}%
\bibitem{BraunMunzinger:2000px}%
  \BibitemOpen
  \bibfield{author}{%
  \bibinfo {author} {\bibfnamefont{P.}~\bibnamefont{Braun-Munzinger}}\ and\
  \bibinfo {author} {\bibfnamefont{J.}~\bibnamefont{Stachel}},\ }%
  \bibfield{journal}{%
  \Doi{10.1016/S0370-2693(00)00991-6}{\bibinfo {journal} {Phys. Lett.}}\ }%
  \textbf{\bibinfo {volume} {B490}},\ \bibinfo {pages} {196} (\bibinfo {year}
  {2000}),\
  \Eprint{http://arxiv.org/abs/nucl-th/0007059}{arXiv:nucl-th/0007059}%
  \bibAnnoteFile{NoStop}{BraunMunzinger:2000px}%
\bibitem{Gorenstein:2000ck}%
  \BibitemOpen
  \bibfield{author}{%
  \bibinfo {author} {\bibfnamefont{M.~I.}\ \bibnamefont{Gorenstein}}, \bibinfo
  {author} {\bibfnamefont{A.~P.}\ \bibnamefont{Kostyuk}}, \bibinfo {author}
  {\bibfnamefont{H.}~\bibnamefont{Stoecker}},\ and\ \bibinfo {author}
  {\bibfnamefont{W.}~\bibnamefont{Greiner}},\ }%
  \bibfield{journal}{%
  \Doi{10.1016/S0370-2693(01)00516-0}{\bibinfo {journal} {Phys. Lett.}}\ }%
  \textbf{\bibinfo {volume} {B509}},\ \bibinfo {pages} {277} (\bibinfo {year}
  {2001}),\ \Eprint{http://arxiv.org/abs/hep-ph/0010148}{arXiv:hep-ph/0010148}%
  \bibAnnoteFile{NoStop}{Gorenstein:2000ck}%
\bibitem{Gorenstein:2000nd}%
  \BibitemOpen
  \bibfield{author}{%
  \bibinfo {author} {\bibfnamefont{M.~I.}\ \bibnamefont{Gorenstein}}, \bibinfo
  {author} {\bibfnamefont{A.}~\bibnamefont{Kostyuk}}, \bibinfo {author}
  {\bibfnamefont{H.}~\bibnamefont{Stoecker}},\ and\ \bibinfo {author}
  {\bibfnamefont{W.}~\bibnamefont{Greiner}},\ }%
  \bibfield{journal}{%
  \Doi{10.1088/0954-3899/27/7/101}{\bibinfo {journal} {J.Phys.G}}\ }%
  \textbf{\bibinfo {volume} {G27}},\ \bibinfo {pages} {L47} (\bibinfo {year}
  {2001}),\ \Eprint{http://arxiv.org/abs/hep-ph/0012015}{arXiv:hep-ph/0012015
  [hep-ph]}%
  \bibAnnoteFile{NoStop}{Gorenstein:2000nd}%
\bibitem{Thews:2000rj}%
  \BibitemOpen
  \bibfield{author}{%
  \bibinfo {author} {\bibfnamefont{R.~L.}\ \bibnamefont{Thews}}, \bibinfo
  {author} {\bibfnamefont{M.}~\bibnamefont{Schroedter}},\ and\ \bibinfo
  {author} {\bibfnamefont{J.}~\bibnamefont{Rafelski}},\ }%
  \bibfield{journal}{%
  \Doi{10.1103/PhysRevC.63.054905}{\bibinfo {journal} {Phys. Rev.}}\ }%
  \textbf{\bibinfo {volume} {C63}},\ \bibinfo {pages} {054905} (\bibinfo {year}
  {2001}),\ \Eprint{http://arxiv.org/abs/hep-ph/0007323}{arXiv:hep-ph/0007323}%
  \bibAnnoteFile{NoStop}{Thews:2000rj}%
\bibitem{Grandchamp:2001pf}%
  \BibitemOpen
  \bibfield{author}{%
  \bibinfo {author} {\bibfnamefont{L.}~\bibnamefont{Grandchamp}}\ and\ \bibinfo
  {author} {\bibfnamefont{R.}~\bibnamefont{Rapp}},\ }%
  \bibfield{journal}{%
  \Doi{10.1016/S0370-2693(01)01311-9}{\bibinfo {journal} {Phys.Lett.}}\ }%
  \textbf{\bibinfo {volume} {B523}},\ \bibinfo {pages} {60} (\bibinfo {year}
  {2001}),\ \Eprint{http://arxiv.org/abs/hep-ph/0103124}{arXiv:hep-ph/0103124
  [hep-ph]}%
  \bibAnnoteFile{NoStop}{Grandchamp:2001pf}%
\bibitem{Zhang:2002ug}%
  \BibitemOpen
  \bibfield{author}{%
  \bibinfo {author} {\bibfnamefont{B.}~\bibnamefont{Zhang}}, \bibinfo {author}
  {\bibfnamefont{C.~M.}\ \bibnamefont{Ko}}, \bibinfo {author}
  {\bibfnamefont{B.-A.}\ \bibnamefont{Li}}, \bibinfo {author}
  {\bibfnamefont{Z.-W.}\ \bibnamefont{Lin}},\ and\ \bibinfo {author}
  {\bibfnamefont{S.}~\bibnamefont{Pal}},\ }%
  \bibfield{journal}{%
  \Doi{10.1103/PhysRevC.65.054909}{\bibinfo {journal} {Phys. Rev.}}\ }%
  \textbf{\bibinfo {volume} {C65}},\ \bibinfo {pages} {054909} (\bibinfo {year}
  {2002}),\
  \Eprint{http://arxiv.org/abs/nucl-th/0201038}{arXiv:nucl-th/0201038}%
  \bibAnnoteFile{NoStop}{Zhang:2002ug}%
\bibitem{Kostyuk:2005zd}%
  \BibitemOpen
  \bibfield{author}{%
  \bibinfo {author} {\bibfnamefont{A.}~\bibnamefont{Kostyuk}}}%
   (\bibinfo {year} {2005}),\
  \Eprint{http://arxiv.org/abs/nucl-th/0502005}{arXiv:nucl-th/0502005
  [nucl-th]}%
  \bibAnnoteFile{NoStop}{Kostyuk:2005zd}%
\bibitem{Yan:2007zze}%
  \BibitemOpen
  \bibfield{author}{%
  \bibinfo {author} {\bibfnamefont{L.}~\bibnamefont{Yan}}, \bibinfo {author}
  {\bibfnamefont{P.}~\bibnamefont{Zhuang}},\ and\ \bibinfo {author}
  {\bibfnamefont{N.}~\bibnamefont{Xu}},\ }%
  \bibfield{journal}{%
  \Doi{10.1142/S0218301307007441}{\bibinfo {journal} {Int. J. Mod. Phys.}}\ }%
  \textbf{\bibinfo {volume} {E16}},\ \bibinfo {pages} {2048} (\bibinfo {year}
  {2007})%
  \bibAnnoteFile{NoStop}{Yan:2007zze}%
\bibitem{Zhao:2007hh}%
  \BibitemOpen
  \bibfield{author}{%
  \bibinfo {author} {\bibfnamefont{X.}~\bibnamefont{Zhao}}\ and\ \bibinfo
  {author} {\bibfnamefont{R.}~\bibnamefont{Rapp}},\ }%
  \bibfield{journal}{%
  \Doi{10.1016/j.physletb.2008.03.068}{\bibinfo {journal} {Phys.Lett.}}\ }%
  \textbf{\bibinfo {volume} {B664}},\ \bibinfo {pages} {253} (\bibinfo {year}
  {2008}),\ \Eprint{http://arxiv.org/abs/0712.2407}{arXiv:0712.2407 [hep-ph]}%
  \bibAnnoteFile{NoStop}{Zhao:2007hh}%
\bibitem{Kostyuk:2003kt}%
  \BibitemOpen
  \bibfield{author}{%
  \bibinfo {author} {\bibfnamefont{A.}~\bibnamefont{Kostyuk}}, \bibinfo
  {author} {\bibfnamefont{M.~I.}\ \bibnamefont{Gorenstein}}, \bibinfo {author}
  {\bibfnamefont{H.}~\bibnamefont{Stoecker}},\ and\ \bibinfo {author}
  {\bibfnamefont{W.}~\bibnamefont{Greiner}},\ }%
  \bibfield{journal}{%
  \Doi{10.1103/PhysRevC.68.041902}{\bibinfo {journal} {Phys.Rev.}}\ }%
  \textbf{\bibinfo {volume} {C68}},\ \bibinfo {pages} {041902} (\bibinfo {year}
  {2003}),\ \Eprint{http://arxiv.org/abs/hep-ph/0305277}{arXiv:hep-ph/0305277
  [hep-ph]}%
  \bibAnnoteFile{NoStop}{Kostyuk:2003kt}%
\bibitem{Adare:2006ns}%
  \BibitemOpen
  \bibfield{author}{%
  \bibinfo {author} {\bibfnamefont{A.}~\bibnamefont{Adare}} \emph{et~al.}
  (\bibinfo {collaboration} {PHENIX}),\ }%
  \bibfield{journal}{%
  \Doi{10.1103/PhysRevLett.98.232301}{\bibinfo {journal} {Phys. Rev. Lett.}}\
  }%
  \textbf{\bibinfo {volume} {98}},\ \bibinfo {pages} {232301} (\bibinfo {year}
  {2007}),\
  \Eprint{http://arxiv.org/abs/nucl-ex/0611020}{arXiv:nucl-ex/0611020}%
  \bibAnnoteFile{NoStop}{Adare:2006ns}%
\bibitem{Adare:2008sh}%
  \BibitemOpen
  \bibfield{author}{%
  \bibinfo {author} {\bibfnamefont{A.}~\bibnamefont{Adare}} \emph{et~al.}
  (\bibinfo {collaboration} {PHENIX}),\ }%
  \bibfield{journal}{%
  \Doi{10.1103/PhysRevLett.101.122301}{\bibinfo {journal} {Phys. Rev. Lett.}}\
  }%
  \textbf{\bibinfo {volume} {101}},\ \bibinfo {pages} {122301} (\bibinfo {year}
  {2008}),\ \Eprint{http://arxiv.org/abs/0801.0220}{arXiv:0801.0220 [nucl-ex]}%
  \bibAnnoteFile{NoStop}{Adare:2008sh}%
\bibitem{Abelev:2009qaa}%
  \BibitemOpen
  \bibfield{author}{%
  \bibinfo {author} {\bibfnamefont{B.}~\bibnamefont{Abelev}} \emph{et~al.}
  (\bibinfo {collaboration} {STAR Collaboration}),\ }%
  \bibfield{journal}{%
  \Doi{10.1103/PhysRevC.80.041902}{\bibinfo {journal} {Phys.Rev.}}\ }%
  \textbf{\bibinfo {volume} {C80}},\ \bibinfo {pages} {041902} (\bibinfo {year}
  {2009}),\ \Eprint{http://arxiv.org/abs/0904.0439}{arXiv:0904.0439 [nucl-ex]}%
  \bibAnnoteFile{NoStop}{Abelev:2009qaa}%
\bibitem{Abelev:2012rv}%
  \BibitemOpen
  \bibfield{author}{%
  \bibinfo {author} {\bibfnamefont{B.}~\bibnamefont{Abelev}} \emph{et~al.}
  (\bibinfo {collaboration} {ALICE})}%
   (\bibinfo {year} {2012}),\
  \Eprint{http://arxiv.org/abs/1202.1383}{arXiv:1202.1383 [hep-ex]}%
  \bibAnnoteFile{NoStop}{Abelev:2012rv}%
\bibitem{Reed:2011fr}%
  \BibitemOpen
  \bibfield{author}{%
  \bibinfo {author} {\bibfnamefont{R.}~\bibnamefont{Reed}},\ }%
  \bibfield{journal}{%
  \Doi{10.1088/0954-3899/38/12/124185}{\bibinfo {journal} {J. Phys.}}\ }%
  \textbf{\bibinfo {volume} {G38}},\ \bibinfo {pages} {124185} (\bibinfo {year}
  {2011}),\ \Eprint{http://arxiv.org/abs/1109.3891}{arXiv:1109.3891 [nucl-ex]}%
  \bibAnnoteFile{NoStop}{Reed:2011fr}%
\bibitem{Chatrchyan:2011pe}%
  \BibitemOpen
  \bibfield{author}{%
  \bibinfo {author} {\bibfnamefont{S.}~\bibnamefont{Chatrchyan}} \emph{et~al.}
  (\bibinfo {collaboration} {CMS Collaboration}),\ }%
  \bibfield{journal}{%
  \Doi{10.1103/PhysRevLett.107.052302}{\bibinfo {journal} {Phys.Rev.Lett.}}\ }%
  \textbf{\bibinfo {volume} {107}},\ \bibinfo {pages} {052302} (\bibinfo {year}
  {2011}),\ \Eprint{http://arxiv.org/abs/1105.4894}{arXiv:1105.4894 [nucl-ex]}%
  \bibAnnoteFile{NoStop}{Chatrchyan:2011pe}%
\bibitem{Abe:1998fb}%
  \BibitemOpen
  \bibfield{author}{%
  \bibinfo {author} {\bibfnamefont{F.}~\bibnamefont{Abe}} \emph{et~al.}
  (\bibinfo {collaboration} {CDF}),\ }%
  \bibfield{journal}{%
  \Doi{10.1103/PhysRevD.58.112004}{\bibinfo {journal} {Phys. Rev.}}\ }%
  \textbf{\bibinfo {volume} {D58}},\ \bibinfo {pages} {112004} (\bibinfo {year}
  {1998}),\ \Eprint{http://arxiv.org/abs/hep-ex/9804014}{arXiv:hep-ex/9804014}%
  \bibAnnoteFile{NoStop}{Abe:1998fb}%
\bibitem{Schroedter:2000ek}%
  \BibitemOpen
  \bibfield{author}{%
  \bibinfo {author} {\bibfnamefont{M.}~\bibnamefont{Schroedter}}, \bibinfo
  {author} {\bibfnamefont{R.~L.}\ \bibnamefont{Thews}},\ and\ \bibinfo {author}
  {\bibfnamefont{J.}~\bibnamefont{Rafelski}},\ }%
  \bibfield{journal}{%
  \Doi{10.1103/PhysRevC.62.024905}{\bibinfo {journal} {Phys. Rev.}}\ }%
  \textbf{\bibinfo {volume} {C62}},\ \bibinfo {pages} {024905} (\bibinfo {year}
  {2000}),\ \Eprint{http://arxiv.org/abs/hep-ph/0004041}{arXiv:hep-ph/0004041}%
  \bibAnnoteFile{NoStop}{Schroedter:2000ek}%
\bibitem{Averbeck:2011ga}%
  \BibitemOpen
  \bibfield{author}{%
  \bibinfo {author} {\bibfnamefont{R.}~\bibnamefont{Averbeck}}, \bibinfo
  {author} {\bibfnamefont{N.}~\bibnamefont{Bastid}}, \bibinfo {author}
  {\bibfnamefont{Z.}~\bibnamefont{del Valle}}, \bibinfo {author}
  {\bibfnamefont{P.}~\bibnamefont{Crochet}}, \bibinfo {author}
  {\bibfnamefont{A.}~\bibnamefont{Dainese}}, \emph{et~al.}}%
   (\bibinfo {year} {2011}),\
  \Eprint{http://arxiv.org/abs/1107.3243}{arXiv:1107.3243 [hep-ph]}%
  \bibAnnoteFile{NoStop}{Averbeck:2011ga}%
\bibitem{ALICE:2011aa}%
  \BibitemOpen
  \bibfield{author}{%
  \bibinfo {author} {\bibfnamefont{B.}~\bibnamefont{Abelev}} \emph{et~al.}
  (\bibinfo {collaboration} {ALICE Collaboration}),\ }%
  \bibfield{journal}{%
  \Doi{10.1007/JHEP01(2012)128}{\bibinfo {journal} {JHEP}}\ }%
  \textbf{\bibinfo {volume} {1201}},\ \bibinfo {pages} {128} (\bibinfo {year}
  {2012}),\ \bibinfo {note} {23 pages, 5 figures},\
  \Eprint{http://arxiv.org/abs/1111.1553}{arXiv:1111.1553 [hep-ex]}%
  \bibAnnoteFile{NoStop}{ALICE:2011aa}%
\bibitem{Emerick:2011xu}%
  \BibitemOpen
  \bibfield{author}{%
  \bibinfo {author} {\bibfnamefont{A.}~\bibnamefont{Emerick}}, \bibinfo
  {author} {\bibfnamefont{X.}~\bibnamefont{Zhao}},\ and\ \bibinfo {author}
  {\bibfnamefont{R.}~\bibnamefont{Rapp}},\ }%
  \bibfield{journal}{%
  \Doi{10.1140/epja/i2012-12072-y}{\bibinfo {journal} {Eur.Phys.J.}}\ }%
  \textbf{\bibinfo {volume} {A48}},\ \bibinfo {pages} {72} (\bibinfo {year}
  {2012}),\ \Eprint{http://arxiv.org/abs/1111.6537}{arXiv:1111.6537 [hep-ph]}%
  \bibAnnoteFile{NoStop}{Emerick:2011xu}%
\bibitem{Nigmanov:2009gu}%
  \BibitemOpen
  \bibfield{author}{%
  \bibinfo {author} {\bibfnamefont{T.}~\bibnamefont{Nigmanov}}, \bibinfo
  {author} {\bibfnamefont{K.}~\bibnamefont{Gibson}}, \bibinfo {author}
  {\bibfnamefont{M.}~\bibnamefont{Hartz}},\ and\ \bibinfo {author}
  {\bibfnamefont{P.}~\bibnamefont{Shepard}} (\bibinfo {collaboration} {CDF
  Collaboration})}%
   (\bibinfo {year} {2009}),\ \bibinfo {note} {to be published in the
  proceedings of DPF-2009, Detroit, MI, July 2009, eConf C090726},\
  \Eprint{http://arxiv.org/abs/0910.3013}{arXiv:0910.3013 [hep-ex]}%
  \bibAnnoteFile{NoStop}{Nigmanov:2009gu}%
\bibitem{Chen:1992fq}%
  \BibitemOpen
  \bibfield{author}{%
  \bibinfo {author} {\bibfnamefont{Y.-Q.}\ \bibnamefont{Chen}}\ and\ \bibinfo
  {author} {\bibfnamefont{Y.-P.}\ \bibnamefont{Kuang}},\ }%
  \bibfield{journal}{%
  \Doi{10.1103/PhysRevD.46.1165}{\bibinfo {journal} {Phys. Rev.}}\ }%
  \textbf{\bibinfo {volume} {D46}},\ \bibinfo {pages} {1165} (\bibinfo {year}
  {1992}),\ \bibinfo {note} {[Erratum-ibid.D47:350,1993]}%
  \bibAnnoteFile{NoStop}{Chen:1992fq}%
\bibitem{Satz:2005hx}%
  \BibitemOpen
  \bibfield{author}{%
  \bibinfo {author} {\bibfnamefont{H.}~\bibnamefont{Satz}},\ }%
  \bibfield{journal}{%
  \Doi{10.1088/0954-3899/32/3/R01}{\bibinfo {journal} {J. Phys.}}\ }%
  \textbf{\bibinfo {volume} {G32}},\ \bibinfo {pages} {R25} (\bibinfo {year}
  {2006}),\ \Eprint{http://arxiv.org/abs/hep-ph/0512217}{arXiv:hep-ph/0512217}%
  \bibAnnoteFile{NoStop}{Satz:2005hx}%
\bibitem{Nakamura:2010zzi}%
  \BibitemOpen
  \bibfield{author}{%
  \bibinfo {author} {\bibfnamefont{K.}~\bibnamefont{Nakamura}} \emph{et~al.}
  (\bibinfo {collaboration} {Particle Data Group}),\ }%
  \bibfield{journal}{%
  \Doi{10.1088/0954-3899/37/7A/075021}{\bibinfo {journal} {J. Phys.}}\ }%
  \textbf{\bibinfo {volume} {G37}},\ \bibinfo {pages} {075021} (\bibinfo {year}
  {2010})%
  \bibAnnoteFile{NoStop}{Nakamura:2010zzi}%
\bibitem{Sjostrand:2000wi}%
  \BibitemOpen
  \bibfield{author}{%
  \bibinfo {author} {\bibfnamefont{T.}~\bibnamefont{Sjostrand}} \emph{et~al.},\
  }%
  \bibfield{journal}{%
  \Doi{10.1016/S0010-4655(00)00236-8}{\bibinfo {journal} {Comput. Phys.
  Commun.}}\ }%
  \textbf{\bibinfo {volume} {135}},\ \bibinfo {pages} {238} (\bibinfo {year}
  {2001}),\ \Eprint{http://arxiv.org/abs/hep-ph/0010017}{arXiv:hep-ph/0010017}%
  \bibAnnoteFile{NoStop}{Sjostrand:2000wi}%
\bibitem{Adare:2010de}%
  \BibitemOpen
  \bibfield{author}{%
  \bibinfo {author} {\bibfnamefont{A.}~\bibnamefont{Adare}} \emph{et~al.}
  (\bibinfo {collaboration} {PHENIX Collaboration}),\ }%
  \bibfield{journal}{%
  \Doi{10.1103/PhysRevC.84.044905}{\bibinfo {journal} {Phys.Rev.}}\ }%
  \textbf{\bibinfo {volume} {C84}},\ \bibinfo {pages} {044905} (\bibinfo {year}
  {2011}),\ \Eprint{http://arxiv.org/abs/1005.1627}{arXiv:1005.1627 [nucl-ex]}%
  \bibAnnoteFile{NoStop}{Adare:2010de}%
\bibitem{Vogt:2010aa}%
  \BibitemOpen
  \bibfield{author}{%
  \bibinfo {author} {\bibfnamefont{R.}~\bibnamefont{Vogt}},\ }%
  \bibfield{journal}{%
  \Doi{10.1103/PhysRevC.81.044903}{\bibinfo {journal} {Phys.Rev.}}\ }%
  \textbf{\bibinfo {volume} {C81}},\ \bibinfo {pages} {044903} (\bibinfo {year}
  {2010}),\ \Eprint{http://arxiv.org/abs/1003.3497}{arXiv:1003.3497 [hep-ph]}%
  \bibAnnoteFile{NoStop}{Vogt:2010aa}%
\bibitem{BraunMunzinger:2001ip}%
  \BibitemOpen
  \bibfield{author}{%
  \bibinfo {author} {\bibfnamefont{P.}~\bibnamefont{Braun-Munzinger}}, \bibinfo
  {author} {\bibfnamefont{D.}~\bibnamefont{Magestro}}, \bibinfo {author}
  {\bibfnamefont{K.}~\bibnamefont{Redlich}},\ and\ \bibinfo {author}
  {\bibfnamefont{J.}~\bibnamefont{Stachel}},\ }%
  \bibfield{journal}{%
  \Doi{10.1016/S0370-2693(01)01069-3}{\bibinfo {journal} {Phys. Lett.}}\ }%
  \textbf{\bibinfo {volume} {B518}},\ \bibinfo {pages} {41} (\bibinfo {year}
  {2001}),\ \Eprint{http://arxiv.org/abs/hep-ph/0105229}{arXiv:hep-ph/0105229}%
  \bibAnnoteFile{NoStop}{BraunMunzinger:2001ip}%
\bibitem{Uphoff:2010sh}%
  \BibitemOpen
  \bibfield{author}{%
  \bibinfo {author} {\bibfnamefont{J.}~\bibnamefont{Uphoff}}, \bibinfo {author}
  {\bibfnamefont{O.}~\bibnamefont{Fochler}}, \bibinfo {author}
  {\bibfnamefont{Z.}~\bibnamefont{Xu}},\ and\ \bibinfo {author}
  {\bibfnamefont{C.}~\bibnamefont{Greiner}},\ }%
  \bibfield{journal}{%
  \Doi{10.1103/PhysRevC.82.044906}{\bibinfo {journal} {Phys.Rev.}}\ }%
  \textbf{\bibinfo {volume} {C82}},\ \bibinfo {pages} {044906} (\bibinfo {year}
  {2010}),\ \Eprint{http://arxiv.org/abs/1003.4200}{arXiv:1003.4200 [hep-ph]}%
  \bibAnnoteFile{NoStop}{Uphoff:2010sh}%
\bibitem{Rafelski:1980gk}%
  \BibitemOpen
  \bibfield{author}{%
  \bibinfo {author} {\bibfnamefont{J.}~\bibnamefont{Rafelski}}\ and\ \bibinfo
  {author} {\bibfnamefont{M.}~\bibnamefont{Danos}},\ }%
  \bibfield{journal}{%
  \Doi{10.1016/0370-2693(80)90601-2}{\bibinfo {journal} {Phys. Lett.}}\ }%
  \textbf{\bibinfo {volume} {B97}},\ \bibinfo {pages} {279} (\bibinfo {year}
  {1980})%
  \bibAnnoteFile{NoStop}{Rafelski:1980gk}%
\bibitem{Andronic:2008gu}%
  \BibitemOpen
  \bibfield{author}{%
  \bibinfo {author} {\bibfnamefont{A.}~\bibnamefont{Andronic}}, \bibinfo
  {author} {\bibfnamefont{P.}~\bibnamefont{Braun-Munzinger}},\ and\ \bibinfo
  {author} {\bibfnamefont{J.}~\bibnamefont{Stachel}},\ }%
  \bibfield{journal}{%
  \Doi{10.1016/j.physletb.2009.02.014}{\bibinfo {journal} {Phys. Lett.}}\ }%
  \textbf{\bibinfo {volume} {B673}},\ \bibinfo {pages} {142} (\bibinfo {year}
  {2009}),\ \bibinfo {note} {[Erratum-ibid.B678:516,2009]},\
  \Eprint{http://arxiv.org/abs/0812.1186}{arXiv:0812.1186 [nucl-th]}%
  \bibAnnoteFile{NoStop}{Andronic:2008gu}%
\bibitem{CMS:2012fr}%
  \BibitemOpen
  \bibfield{author}{%
  \bibinfo {author} {\bibfnamefont{S.}~\bibnamefont{Chatrchyan}} \emph{et~al.}
  (\bibinfo {collaboration} {CMS Collaboration})}%
   (\bibinfo {year} {2012}),\
  \Eprint{http://arxiv.org/abs/1208.2826}{arXiv:1208.2826 [nucl-ex]}%
  \bibAnnoteFile{NoStop}{CMS:2012fr}%
\bibitem{Peskin:1979va}%
  \BibitemOpen
  \bibfield{author}{%
  \bibinfo {author} {\bibfnamefont{M.~E.}\ \bibnamefont{Peskin}},\ }%
  \bibfield{journal}{%
  \Doi{10.1016/0550-3213(79)90199-8}{\bibinfo {journal} {Nucl. Phys.}}\ }%
  \textbf{\bibinfo {volume} {B156}},\ \bibinfo {pages} {365} (\bibinfo {year}
  {1979})%
  \bibAnnoteFile{NoStop}{Peskin:1979va}%
\bibitem{Bhanot:1979vb}%
  \BibitemOpen
  \bibfield{author}{%
  \bibinfo {author} {\bibfnamefont{G.}~\bibnamefont{Bhanot}}\ and\ \bibinfo
  {author} {\bibfnamefont{M.~E.}\ \bibnamefont{Peskin}},\ }%
  \bibfield{journal}{%
  \Doi{10.1016/0550-3213(79)90200-1}{\bibinfo {journal} {Nucl.Phys.}}\ }%
  \textbf{\bibinfo {volume} {B156}},\ \bibinfo {pages} {391} (\bibinfo {year}
  {1979})%
  \bibAnnoteFile{NoStop}{Bhanot:1979vb}%
\bibitem{Arleo:2001mp}%
  \BibitemOpen
  \bibfield{author}{%
  \bibinfo {author} {\bibfnamefont{F.}~\bibnamefont{Arleo}}, \bibinfo {author}
  {\bibfnamefont{P.~B.}\ \bibnamefont{Gossiaux}}, \bibinfo {author}
  {\bibfnamefont{T.}~\bibnamefont{Gousset}},\ and\ \bibinfo {author}
  {\bibfnamefont{J.}~\bibnamefont{Aichelin}},\ }%
  \bibfield{journal}{%
  \Doi{10.1103/PhysRevD.65.014005}{\bibinfo {journal} {Phys. Rev.}}\ }%
  \textbf{\bibinfo {volume} {D65}},\ \bibinfo {pages} {014005} (\bibinfo {year}
  {2002}),\ \Eprint{http://arxiv.org/abs/hep-ph/0102095}{arXiv:hep-ph/0102095}%
  \bibAnnoteFile{NoStop}{Arleo:2001mp}%
\bibitem{Polleri:2003kn}%
  \BibitemOpen
  \bibfield{author}{%
  \bibinfo {author} {\bibfnamefont{A.}~\bibnamefont{Polleri}}, \bibinfo
  {author} {\bibfnamefont{T.}~\bibnamefont{Renk}}, \bibinfo {author}
  {\bibfnamefont{R.}~\bibnamefont{Schneider}},\ and\ \bibinfo {author}
  {\bibfnamefont{W.}~\bibnamefont{Weise}},\ }%
  \bibfield{journal}{%
  \Doi{10.1103/PhysRevC.70.044906}{\bibinfo {journal} {Phys. Rev.}}\ }%
  \textbf{\bibinfo {volume} {C70}},\ \bibinfo {pages} {044906} (\bibinfo {year}
  {2004}),\
  \Eprint{http://arxiv.org/abs/nucl-th/0306025}{arXiv:nucl-th/0306025}%
  \bibAnnoteFile{NoStop}{Polleri:2003kn}%
\bibitem{Kaczmarek:2005ui}%
  \BibitemOpen
  \bibfield{author}{%
  \bibinfo {author} {\bibfnamefont{O.}~\bibnamefont{Kaczmarek}}\ and\ \bibinfo
  {author} {\bibfnamefont{F.}~\bibnamefont{Zantow}},\ }%
  \bibfield{journal}{%
  \Doi{10.1103/PhysRevD.71.114510}{\bibinfo {journal} {Phys. Rev.}}\ }%
  \textbf{\bibinfo {volume} {D71}},\ \bibinfo {pages} {114510} (\bibinfo {year}
  {2005}),\
  \Eprint{http://arxiv.org/abs/hep-lat/0503017}{arXiv:hep-lat/0503017}%
  \bibAnnoteFile{NoStop}{Kaczmarek:2005ui}%
\bibitem{Shuryak:2004tx}%
  \BibitemOpen
  \bibfield{author}{%
  \bibinfo {author} {\bibfnamefont{E.~V.}\ \bibnamefont{Shuryak}}\ and\
  \bibinfo {author} {\bibfnamefont{I.}~\bibnamefont{Zahed}},\ }%
  \bibfield{journal}{%
  \Doi{10.1103/PhysRevD.70.054507}{\bibinfo {journal} {Phys. Rev.}}\ }%
  \textbf{\bibinfo {volume} {D70}},\ \bibinfo {pages} {054507} (\bibinfo {year}
  {2004}),\ \Eprint{http://arxiv.org/abs/hep-ph/0403127}{arXiv:hep-ph/0403127}%
  \bibAnnoteFile{NoStop}{Shuryak:2004tx}%
\bibitem{Kolb:2000fha}%
  \BibitemOpen
  \bibfield{author}{%
  \bibinfo {author} {\bibfnamefont{P.}~\bibnamefont{Kolb}}, \bibinfo {author}
  {\bibfnamefont{P.}~\bibnamefont{Huovinen}}, \bibinfo {author}
  {\bibfnamefont{U.~W.}\ \bibnamefont{Heinz}},\ and\ \bibinfo {author}
  {\bibfnamefont{H.}~\bibnamefont{Heiselberg}},\ }%
  \bibfield{journal}{%
  \Doi{10.1016/S0370-2693(01)00079-X}{\bibinfo {journal} {Phys.Lett.}}\ }%
  \textbf{\bibinfo {volume} {B500}},\ \bibinfo {pages} {232} (\bibinfo {year}
  {2001}),\ \Eprint{http://arxiv.org/abs/hep-ph/0012137}{arXiv:hep-ph/0012137
  [hep-ph]}%
  \bibAnnoteFile{NoStop}{Kolb:2000fha}%
\bibitem{Zhu:2004nw}%
  \BibitemOpen
  \bibfield{author}{%
  \bibinfo {author} {\bibfnamefont{X.}~\bibnamefont{Zhu}}, \bibinfo {author}
  {\bibfnamefont{P.}~\bibnamefont{Zhuang}},\ and\ \bibinfo {author}
  {\bibfnamefont{N.}~\bibnamefont{Xu}},\ }%
  \bibfield{journal}{%
  \Doi{10.1016/j.physletb.2004.12.023}{\bibinfo {journal} {Phys. Lett.}}\ }%
  \textbf{\bibinfo {volume} {B607}},\ \bibinfo {pages} {107} (\bibinfo {year}
  {2005}),\
  \Eprint{http://arxiv.org/abs/nucl-th/0411093}{arXiv:nucl-th/0411093}%
  \bibAnnoteFile{NoStop}{Zhu:2004nw}%
\bibitem{Chatrchyan:2012np}%
  \BibitemOpen
  \bibfield{author}{%
  \bibinfo {author} {\bibfnamefont{S.}~\bibnamefont{Chatrchyan}} \emph{et~al.}
  (\bibinfo {collaboration} {CMS Collaboration})}%
   (\bibinfo {year} {2012}),\
  \Eprint{http://arxiv.org/abs/1201.5069}{arXiv:1201.5069 [nucl-ex]}%
  \bibAnnoteFile{NoStop}{Chatrchyan:2012np}%
\bibitem{Liu:2009wza}%
  \BibitemOpen
  \bibfield{author}{%
  \bibinfo {author} {\bibfnamefont{Y.}~\bibnamefont{Liu}}, \bibinfo {author}
  {\bibfnamefont{Z.}~\bibnamefont{Qu}}, \bibinfo {author}
  {\bibfnamefont{N.}~\bibnamefont{Xu}},\ and\ \bibinfo {author}
  {\bibfnamefont{P.}~\bibnamefont{Zhuang}},\ }%
  \bibfield{journal}{%
  \Doi{10.1088/0954-3899/37/7/075110}{\bibinfo {journal} {J. Phys.}}\ }%
  \textbf{\bibinfo {volume} {G37}},\ \bibinfo {pages} {075110} (\bibinfo {year}
  {2010}),\ \Eprint{http://arxiv.org/abs/0907.2723}{arXiv:0907.2723 [nucl-th]}%
  \bibAnnoteFile{NoStop}{Liu:2009wza}%
\bibitem{Aamodt:2010cz}%
  \BibitemOpen
  \bibfield{author}{%
  \bibinfo {author} {\bibfnamefont{K.}~\bibnamefont{Aamodt}} \emph{et~al.}
  (\bibinfo {collaboration} {ALICE Collaboration}),\ }%
  \bibfield{journal}{%
  \bibinfo {journal} {Phys.Rev.Lett.}\ }%
  \textbf{\bibinfo {volume} {106}},\ \bibinfo {pages} {032301} (\bibinfo {year}
  {2011}),\ \Eprint{http://arxiv.org/abs/1012.1657}{arXiv:1012.1657 [nucl-ex]}%
  \bibAnnoteFile{NoStop}{Aamodt:2010cz}%
\bibitem{Kuznetsova:2006bh}%
  \BibitemOpen
  \bibfield{author}{%
  \bibinfo {author} {\bibfnamefont{I.}~\bibnamefont{Kuznetsova}}\ and\ \bibinfo
  {author} {\bibfnamefont{J.}~\bibnamefont{Rafelski}},\ }%
  \bibfield{journal}{%
  \Doi{10.1140/epjc/s10052-007-0268-9}{\bibinfo {journal} {Eur.Phys.J.}}\ }%
  \textbf{\bibinfo {volume} {C51}},\ \bibinfo {pages} {113} (\bibinfo {year}
  {2007}),\ \Eprint{http://arxiv.org/abs/hep-ph/0607203}{arXiv:hep-ph/0607203
  [hep-ph]}%
  \bibAnnoteFile{NoStop}{Kuznetsova:2006bh}%
\bibitem{Liu:2009nb}%
  \BibitemOpen
  \bibfield{author}{%
  \bibinfo {author} {\bibfnamefont{Y.}~\bibnamefont{Liu}}, \bibinfo {author}
  {\bibfnamefont{Z.}~\bibnamefont{Qu}}, \bibinfo {author}
  {\bibfnamefont{N.}~\bibnamefont{Xu}},\ and\ \bibinfo {author}
  {\bibfnamefont{P.}~\bibnamefont{Zhuang}},\ }%
  \bibfield{journal}{%
  \Doi{10.1016/j.physletb.2009.06.006}{\bibinfo {journal} {Phys. Lett.}}\ }%
  \textbf{\bibinfo {volume} {B678}},\ \bibinfo {pages} {72} (\bibinfo {year}
  {2009}),\ \Eprint{http://arxiv.org/abs/0901.2757}{arXiv:0901.2757 [nucl-th]}%
  \bibAnnoteFile{NoStop}{Liu:2009nb}%
\end{thebibliography}%

\end{document}